      [1999/12/01 v1.4c Il Nuovo Cimento]
\documentclass{cimento}

\title{Stationary spacetimes and the Simon  tensor}
\author{D.~Bini\from{ins:x}\from{ins:y} and R. T. Jantzen\from{ins:z}\from{ins:y}}

\instlist{
\inst{ins:x}{Istituto per le Applicazioni del Calcolo \lq\lq M. Picone\rq\rq, C.N.R.,
   I-- 00161 Roma, Italy
} 
\inst{ins:y} {International Center for Relativistic Astrophysics, I.C.R.A., 
University of Rome ``La Sapienza,'' I--00185 Rome, Italy.}

\inst{ins:z}{Department of Mathematical Sciences, Villanova University,
  Villanova, PA 19085, USA} 
}

\def\beq{\begin{equation}}
\def\eeq{\end{equation}}

\def\dualp#1{{}^{\ast_{(\hbox{$\scriptstyle #1$})}} \kern-1pt}
\def\dual{\,{}^\ast{}\kern-1.5pt}

  \def\del{\nabla}
\def\div{\mathop{\rm div}\nolimits}    \def\curl{\mathop{\rm curl}\nolimits}
\def\Scurl{\mathop{\rm Scurl}\nolimits} 
\def\TF{{}^{(\rm TF)}}
\def\SYM{\mathop{\rm SYM}\nolimits}  

\def\rmd{{\rm d}}
\def\fl{}

\begin{document}

\maketitle

\begin{abstract}
For stationary vacuum spacetimes the Bianchi identities of the second kind  equate  
the Simon tensor to the Simon-Mars tensor, the latter having a clear geometrical interpretation. The equivalence of these two tensors is broken in the nonvacuum case by additional source energy-momentum terms, but
absorbing these source terms into a redefinition of the Simon tensor restores the equality.
Explicit examples are discussed for electrovacuum and rigidly rotating matter fields.
\end{abstract}

\section{Introduction}
Using the $1+3$ splitting of spacetime associated with an observer congruence $u$ (timelike, unit tangent vector to the congruence), 
the Bianchi identities for the curvature tensor can be cast into a Maxwell-like form for the electric and magnetic parts of the Weyl tensor with sources arising from the Ricci tensor terms, in turn re-expressed in terms of the energy-momentum tensor of the spacetime through the Einstein field equations \cite{Ellis73,Maartens,Ellis99}.

For the stationary case with $u$ directed along the timelike Killing vector field, these equations are greatly simplified and suggested the introduction of the (complex) Simon \cite{Simon} and  Simon-Mars tensors \cite{Mars99,Mars00,Mars01,Fasop99, Fasop00,cys1,cys2}. 
The Simon tensor  was introduced for vacuum stationary spacetimes as a complex generalization of 
the Cotton-York tensor \cite{Cotton,York}, to which it reduces in the static case.
The Simon-Mars tensor, instead, is a natural algebraic combination of the self-dual fields associated with the Papapetrou field of the timelike Killing vector field and the Weyl tensor. 
In the vacuum case the Bianchi identities state the equivalence between these two tensors, and
the vanishing of the Simon-Mars tensor has been understood as corresponding to the  alignment of the principal null directions of the Weyl tensor of the spacetime with those of the Papapetrou field associated with the timelike Killing congruence. This is exactly what happens in the case of the Kerr solution, uniquely characterized by the vanishing of the Simon (and Simon-Mars) tensor. 

In the nonvacuum case the equivalence between the Simon and the Simon-Mars tensor is broken because of the presence of Ricci tensor terms in the Bianchi identities which may be re-expressed in terms of the source energy-momentum tensor by the Einstein equations.  
To characterize exact solutions in a way similar way to the vacuum case, one approach would be to study these distinct tensors to explore their meaning.
Alternatively, one can absorb the source terms into a redefinition of the Simon tensor so that the resulting generalized Simon tensor equals the Simon-Mars tensor, which continues to have a clear geometrical meaning. In this more geometrically motivated case,  one can classify exact solutions studying only the vanishing of a single tensor.
Here we apply these considerations to
electrovac stationary spacetimes which contain a source-free electromagnetic field \cite{cys2}
and stationary  spacetimes containing a rigidly rotating perfect fluid\cite{Perjes}.
 
\section{The $1+3$ Maxwell-like equations for the Weyl tensor with sources}

The Bianchi identities of the second kind equate the divergence of the Weyl curvature tensor $C_{\alpha\beta\gamma\delta}$ to half the Cotton tensor \cite{Cotton}, which can be thought of as a current in analogy with electromagnetism \cite{HE}
(where $F^{\alpha\beta}{}_{;\beta} = 4\pi J^\alpha$)
\beq
\label{current}
 C^{\alpha\delta}{}_{\beta\gamma;\delta}
=  -\del_{[\beta} ( {}^{(4)}R^\alpha{}_{\gamma ]}  -\frac16 {}^{(4)}R \delta^\alpha{}_{\gamma ]} )
\equiv J^\alpha{}_{\beta\gamma}\ ,
\eeq
following the conventions of Misner, Thorne and Wheeler \cite{mtw}, including the signature $+2$.
It is convenient to single out the Ricci tensor (RT) and the Ricci scalar (RS) contributions to the Cotton current  $J^\alpha{}_{\beta\gamma}$ of Eq.~(\ref{current}):
\beq
J^\alpha{}_{\beta\gamma} = J_{\rm (RT)}{}^\alpha{}_{\beta\gamma}+J_{\rm (RS)}{}^\alpha{}_{\beta\gamma},
\eeq
where
\beq\fl\qquad
J_{\rm (RT)}{}^\alpha{}_{\beta\gamma}= -\del_{[\beta} ( {}^{(4)}R^\alpha{}_{\gamma ]} ),\qquad
J_{\rm (RS)}{}^\alpha{}_{\beta\gamma}= \frac16 \del_{[ \beta} {}^{(4)}R \delta^\alpha{}_{\gamma ]} .
\eeq
The Einstein equations (in Ricci form) 
\beq\label{eq:EEricci}
  R^\alpha{}_\beta 
 = \kappa(T^\alpha{}_\beta - {\textstyle\frac12} T^\gamma{}_\gamma\delta^\alpha{}_\beta)
\eeq
can then be used to replace the Ricci tensor and scalar curvature terms in this current by the energy-momentum tensor:
\beq
\label{current_T}
 J^\alpha{}_{\beta\gamma}=  -\kappa \del_{[\beta} ( T^\alpha{}_{\gamma ]}  -\frac13 T \delta^\alpha{}_{\gamma ]} ).
\eeq
Splitting these equations with respect to a generic
timelike congruence $u$ leads to their $1+3$ Maxwell-like form  \cite{Ellis99}
given explicitly in our notation in \cite{cys1, cys2}.
The Weyl tensor splits into two symmetric tracefree spatial fields,  its electric and magnetic parts respectively
\beq
\fl\qquad
  E \,^{\alpha} {}_{\beta} 
=  C^{\alpha} {}_{\gamma\beta\delta}\, u^\gamma \, u^\delta \ ,\quad
  H \,^{\alpha} {}_{\beta} 
= - {}^\ast C^{\alpha} {}_{\gamma\beta\delta}\, u^\gamma \, u^\delta
= \frac{1}{2}\, \eta \, ^{\alpha} {}_{\gamma} {}^{\delta}\,
 C^{\gamma}{}_{\delta\beta\rho}\, u^\rho 
\ .
\eeq
If $a$ is the acceleration vector and $\omega$ the vorticity vector of $u$,
introducing the complex spatial fields from the gravitoelectromagnetic connection and Weyl curvature fields
\beq
    z=-a-i\omega = g-i\vec H/2 \ ,\qquad
    Z=E-iH\ ,
\eeq
where $g$ and $\vec H$ are the gravitoelectric and gravitomagnetic vector fields,
leads to a more efficient representation of those equations.
Here we specialize to the case of a stationary spacetime
and align $u = M^{-1}\xi$ with the associated Killing vector $\xi$, so that
the acceleration can be expressed as $a = \nabla \ln M$, where 
$M=|-\xi_\alpha \xi^\alpha|^{1/2}$. 

Splitting the Weyl divergence equations and expressing them in terms of an adapted frame 
$\{e_\top =u, e_a\}$, $a=1,2,3$, where the spatial frame $\{e_a\}$ is orthogonal to $u$, 
one finds in an index-free notation the identities \cite{cys1}
\beq\fl\qquad
\label{bianchi}
M^{3}\div [M^{-3} Z] = 3 z \cdot Z +  \rho^{(G)}
\ ,\quad
M^{-1}\Scurl [MZ] =  z\times Z + i J^{(G)}
\ ,\eeq
where the complex current fields are defined by
\beq
\fl\qquad
\rho^{(G)}_a=J^\top{}_{a\top}+iJ^{*} {}^\top{}_{a\top}\ , \qquad 
J^{(G)}_{ab}=-[J_{(ab)\top}+i J^{*}{}_{(ab)\top}] \ .
\eeq
The div and Scurl spatial operators on symmetric spatial 2-tensors $S$ 
\beq
  [\div(u) S]^\alpha = \del(u)_\beta S^{\alpha\beta}\ ,\
  [\Scurl(u) S]^{\alpha\beta} 
     = \eta(u)^{\gamma\delta(\alpha} \nabla(u)_\gamma S^{\beta)}{}_\delta
\eeq
are defined in terms of the
spatial covariant derivative $\nabla(u)$, the spatial volume 3-form $\eta(u)_{\alpha\beta\gamma} 
=u^\delta \eta_{\delta\alpha\beta\gamma}$ 
(associated with the spatial duality operation $\dualp{u}$, used for example to define the vorticity vector $\omega(u)^\alpha=\frac12\eta(u)^{\alpha\beta\gamma}\nabla_\beta u_\gamma$ from the corresponding 2-form) and the spatial cross and dot products
associated with $u$, 
all given explicitly in \cite{cys1} by spatially projecting spacetime derivatives and quantities orthogonally to $u$. 
For example, the cross-product of a spatial vector and spatial symmetric tensor is
\beq
  [X \times_u A]^{\alpha\beta} = \eta (u)^{\,\gamma\delta (\alpha }
  X_{\gamma} A^{\beta)} {}_{\delta}\ .
\eeq
For simplicity we drop the reference to $u$ on these spatially projected operators, resorting to the abbreviation $\vec\nabla=\nabla(u)$ to distinguish the spatial covariant derivative from the spacetime covariant derivative $\nabla$,
an unnecessary distinction when acting on stationary scalars. In an adapted frame spatial quantities may be expressed using only Latin indexed components.

In the Scurl equation in the projected Bianchi identity (\ref{bianchi}) we have three terms:
\begin{enumerate}
\item
The (rescaled) Scurl of $Z$ which defines the complex, symmetric and tracefree Simon 2-tensor \cite{Simon,cys1} of the vacuum spacetime
\beq
\label{SIMON}
{\rm Simon(vacuum)} =M^{-1}\Scurl [MZ].
\eeq

\item 
The spatial cross product of $z$ and $Z$  which defines the complex, symmetric and tracefree Simon-Mars 2-tensor \cite{Mars99,Mars00,Mars01,cys1}
\beq
\label{simon}
{\rm SimonMars} = z\times Z\ .
\eeq
\item
The Cotton gravitational current term $iJ^{(G)}$, which is also complex, symmetric and tracefree. Its expression in terms of Ricci tensor components is
\begin{eqnarray}
\label{JGfromdef}
\fl\qquad
J^{(G)}_{ab}&=&
\frac{1}{2M^2}\nabla_{(a}M^2{}^{(4)}R_{b)\top} 
-\frac12 (\omega \times {}^{(4)}{\rm Ricci})_{ab} \ ,
\nonumber \\
\fl\qquad
&&-\frac{i}{2} ({\rm Scurl }\, {}^{(4)}{\rm Ricci})_{ab} 
+\frac{3i}{2}[\omega_{(a}{}^{(4)}R_{ b)\top}]^{\rm (TF)} \ .
\end{eqnarray}
\end{enumerate}

Here ${}^{(4)}{\rm Ricci}$ denotes the index-free form  of ${}^{(4)}R_{ab}$ to distinguish it from the spacetime curvature scalar ${}^{(4)}R$.
The only nonvanishing field arising from splitting the Ricci scalar contribution to the Cotton current is the spatial tensor
\beq
J_{\rm  (RS)}{}^a{}_{bc}=\frac{1}{6}\, \partial_{[b}{}^{(4)} R\delta^a_{c]}\ ,
\eeq
so that
$$
J_{\rm (RS)}{}^a{}_{b\top}=0\ , \qquad 
{}^*J_{\rm (RS)}{}^a{}_{b\top} = -\frac{1}{12}\eta^a{}_{b}{}^c \partial_c {}^{(4)}R\ .
$$
Thus ${}^*J_{\rm (RS)}{}_{(ab)\top}=0$
and hence it does not contribute at all to $J^{(G)}$.

Let us define the nonvacuum Simon 2-tensor for any stationary spacetime (with any energy-momentum source) as
\beq
  {\rm Simon(nonvacuum)} = {\rm Simon(vacuum)} - i J^{(G)}\ ,
\eeq
so that the Simon(nonvacuum), hereafter denoted only by Simon, and the Simon-Mars tensors also remain equal in the nonvacuum case. Of course once the Ricci tensor in the source term $J^{(G)}$ is replaced using the Einstein equations (\ref{eq:EEricci}), this equality between Simon and Simon-Mars is only true if they are each evaluated on a solution of those equations, in contrast to the vacuum case where they are identically equal for any metric.
With this understanding there is no longer any need to use two different names, Simon and Simon-Mars, to denote the same quantity.

It is worth noting further that this identification has a true geometrical content. In fact,
the vanishing of the Simon-Mars tensor ($z\times Z=0$) is equivalent to
the alignment of $z$ and $Z$, in the sense that $Z$ must be proportional to the tracefree tensor product of $z$ with itself \cite{cys1}
\beq
Z \propto \, [z\otimes z]^{\rm(TF)} \ .
\eeq
Of course, other generalizations of Simon(vacuum) can also be studied. 

Finally, we remark that when working with explicit solutions of the Einstein field equations, $J^{(G)}$ can have  useful special forms.
For example, when $J^{(G)} = -i \nabla \ln\sigma \times Z$, then the Scurl identity for any scalar $\sigma$ and any symmetric spatial 2-tensor $A$
\beq
\sigma\Scurl [\sigma^{-1} A] =-\vec\nabla \ln\sigma \times A + \Scurl A 
\eeq
enables this current to be incorporated into the existing Scurl term
\beq\label{eq:simonsigma}
{\rm Simon} = (\sigma/M)\Scurl [\sigma^{-1} M Z]  \ .
\eeq
This again would re-establish the vacuum property that the Simon tensor is proportional to a Scurl. This is the case with the family of Kerr-Newman-Taub-NUT electrovac spacetimes (including the more physically interesting Kerr-Newman spacetime), where the extra factor of $\sigma$ converts this to the same Scurl expression when written in terms of the Ernst potential as in the vacuum formula (modulo a scale factor).
For rigidly rotating perfect fluid solutions, as in the case of the Kramer solution \cite{Kramersol}, $J^{(G)}$ is exactly zero and the vacuum case formulas again hold.

\section{Electrovac spacetimes: Maxwell equations and the Papapetrou fields}

The stationary Maxwell-like equations (\ref{bianchi}) for the Weyl tensor expressed in complex self-dual form 
are very similar to the stationary  Maxwell equations expressed in complex form \cite{cys1}
\beq
\fl\qquad
\label{eq:max}
M^{-2}\div(M^2 x) = -2 z\cdot x+4\pi \rho \ ,\qquad
M^{-1}\curl(M x)=4\pi i J
\eeq
satisfied by the spacetime's electromagnetic field (using an arrow to distinguish the electric field
from the electric part of the Weyl tensor in index-free notation) 
\beq
 F=u\wedge \vec E + \dualp{u} B\ ,\qquad x = \vec E -iB\ .
\eeq
They only differ in form by the missing $z\times x$ term in the curl equation here, where $\dualp{u}B$ is the spatial dual of the magnetic 1-form field $B$.
However, when $z$ and $Z$ are aligned (i.e., $z\times Z=0$), the correspondence is complete. The sources $\rho$ and $J$ for the spacetime electromagnetic field are zero for the electrovac spacetimes considered here.
The curl equation for the electromagnetic field implies that $Mx$ is the gradient of a potential
\beq\label{potx}
Mx=\nabla \Phi\ , \quad{\rm or}\quad  x=M^{-1}\nabla \Phi\ .
\eeq 

Papapetrou \cite{pap66} observed that a Killing vector field $\xi$ can be
interpreted as a Lorentz gauge vector potential for an electromagnetic field later called the Papapetrou field \cite{Fasop99,Fasop00}
\beq\label{eq:papfield}
\widetilde F_{\alpha\beta}
=[d \xi]_{\alpha\beta}
= \nabla_\alpha \xi_\beta - \nabla_\beta \xi_\alpha
= 2\nabla_\alpha \xi_\beta 
\eeq
satisfying Maxwell's equations with an effective current source proportional to the Ricci tensor
\beq\label{eq:papst}
\fl\qquad
 \xi_{(\alpha;\beta)}=0 
\rightarrow
 \xi^\alpha{}_{;\alpha}=0 
\rightarrow
 \xi_{\alpha;\beta}{}^{;\beta} = -\xi_\beta R^\beta{}_\alpha 
\rightarrow
\tilde F_{\alpha\beta}{}^{;\beta} = 2\xi_\beta R^\beta{}_\alpha \ ,
\eeq
where this divergence condition is merely the trace of the Ricci identity for a Killing vector field.
Rescaling the Papapetrou field $\tilde F$ by the factor $(2M)^{-1}$ \cite{cys1}
and introducing its complex self-dual representation yields the 
self-dual 2-form $\mathcal{F}$ associated with $z$
 \beq\label{eq:Papselfdual}
\mathcal{F} = (2M)^{-1}(\tilde F + i \dual \tilde F) 
  = u\wedge z +i\dualp{u} z \ ,
\eeq
which when multiplied by $M$
satisfies the same Maxwell equations but with nonzero sources generated by the electromagnetic field. These equations
\beq
\label{eqpap}
\fl\qquad
 M^{-3}\div(M^3z)=-2 z\cdot z- x\cdot \bar x\ ,\ \quad
 M^{-2} \curl(M^2 z) =  x \times \bar x \ .
\eeq
are obtained by splitting the vector Eqs.~(\ref{eq:papst}) and
using the Einstein equations (\ref{eq:EEricci}) to replace the Ricci tensor appearing there in terms of the appropriate projections of the electromagnetic energy-momentum tensor of the spacetime
\beq
\kappa T^{\top}{}_{\top} = - x\cdot \bar x\ ,\quad
 i \kappa T^{\top}{}_{a} =  [x \times \bar x]_a \ ,
\eeq
which then leads to the source terms appearing in these Maxwell equations.

Inserting the potential representation (\ref{potx}) of $Mx$ into the rescaled Papapetrou field curl equation of Eqs.~(\ref{eqpap}) then yields
\beq
M^{-2} \curl(M^2 z) = M^{-2} \nabla \Phi \times \nabla \bar\Phi =- M^{-2} \curl (\bar \Phi \nabla \Phi ) \ ,
\eeq
namely 
\beq\label{eq:MPhi}
\curl(M^2 z + \bar \Phi \nabla \Phi)=0\ .
\eeq
This latter quantity in parentheses must therefore be a gradient, defining
the Ernst potential $\mathcal{E}$
\cite{Kramer} (see their Eq.~(18.34) with $F=-M^2$, reducing to the integrated Eq.~(18.29) in the vacuum case $\Phi=0$) modulo a numerical factor
\beq\label{eq:ernstpot}
\nabla \mathcal{E} = -2 (M^2 z + \bar \Phi \nabla \Phi) \ .
\eeq

The remaining divergence equation of Eqs.~(\ref{eqpap}) is the Ernst equation, which can be expressed in the form (equivalent to Eq.~(18.38) of \cite{Kramer}, see Appendix A)
\beq
\label{ernst}
M \div (M\nabla \mathcal{E} ) = - 2M^2 z\cdot \nabla \mathcal{E}\ .
\eeq
This result follows from 
using the definition of $\mathcal{E}$ to  re-express $x$
in terms of the gradient of $\Phi$ in the two divergence equations (\ref{eq:max}) and (\ref{eqpap}) 
\beq
\fl\qquad
\frac12 \div \nabla \mathcal{E} +\bar \Phi \div \nabla \Phi = M^2 (a+2z)\cdot z, \qquad
\div \nabla \Phi = - (a+2z)\cdot \nabla \Phi
\eeq
and using the second of these equations to eliminate $\div \nabla \Phi$ in the first, yielding
\beq
\div \nabla \mathcal{E} +a \cdot \nabla \mathcal{E}= - 2z\cdot \nabla \mathcal{E}\ ,
\eeq
which can then be rewritten as Eq.~(\ref{ernst}).

The Simon-Mars tensor can be evaluated in a form which shows its relation to the Ernst potential expression for the Simon tensor, assuming that the Einstein field equations are satisfied so that the two tensors are in fact equal.
Following the notation of \cite{cys1},
the purely spatial components and the mixed time-space components of the Ricci tensor field equations (see \cite{Kramer}, Eqs.~(18.15) and (18.16)) are respectively
\begin{eqnarray}
\label{ricci}
[{}^{(4)} {\rm Ricci}]^{\rm (TF)}_{ab}
&=& -2M^{-2}[\nabla_{(a}\Phi \nabla_{b)}\bar\Phi]^{\rm (TF)}\ ,\nonumber \\
{}^{(4)}R_{\top b} 
&=& iM^{-2}[\curl (M^2 z)]_b 
   =-iM^{-2}[\curl (\bar \Phi \nabla \Phi)]_b \ ,
\end{eqnarray}
where $z$ and $\Phi$ are related to the Ernst potential by (\ref{eq:ernstpot}).
A straightforward computation gives the electric and magnetic part of the Weyl tensor in terms of the gravitoelectromagnetic fields $a$ and $\omega$ and the Ricci tensor of the quotient 3-manifold \cite{cys1}
\begin{eqnarray}
E&=& -\frac12 [4 \omega \otimes \omega -\vec\nabla a -a\otimes a -^{(3)}{\rm Ricci}]^{\rm (TF)}\ , \nonumber \\
H&=&-\SYM [\vec\nabla \omega +2 a \otimes \omega]^{\rm (TF)}\ .
\end{eqnarray}
Then using the formula
\beq
\label{R3vsR4}
[{}^{(3)}{\rm Ricci}]^{\rm (TF)}
=[{}^{(4)}{\rm Ricci} + \vec\nabla a + a\otimes a +2 \omega \otimes \omega]^{\rm (TF)}\ ,
\eeq
one finds
\beq
\label{Zdef}
Z = E-iH
  =[-\SYM(\vec\nabla z) + z \otimes z + \frac12 {}^{(4)}{\rm Ricci}]^{\rm (TF)}\ .
\eeq
The Simon-Mars tensor is then
\beq
{\rm SimonMars}
  = z\times (-\SYM \,\vec\nabla z+\frac12 {}^{(4)}{\rm Ricci})
\eeq
since $z\times(z\otimes z)$ is identically zero. 
By next replacing the first term here using the Scurl identity valid for an arbitrary spatial vector field $X$
\beq
\fl\qquad
\Scurl  (X\otimes X) 
= -X\times \SYM  (\vec\nabla X) + \frac32 [\SYM (X\otimes \curl X)]^{\rm (TF)}\ ,
\eeq
and using the Einstein Eqs.~(\ref{ricci}) to replace the spatial components of the spacetime Ricci tensor, the Simon-Mars tensor becomes
\begin{eqnarray}
\label{nuovosimon}
{\rm SimonMars}
&=& \Scurl (z\otimes z) -\frac32 [\SYM (z\otimes \curl \,z)]^{\rm (TF)}
\nonumber\\ &&\quad
 -M^{-2} z\times  \SYM (\nabla \Phi \otimes \nabla\bar\Phi )\ .
\end{eqnarray}
Finally by substituting $z=M^{-2}(M^2 z)$ into the curl term here expanding the result with the appropriate product rule and incorporating the resulting additional acceleration term into the Scurl term, and then using Eq.~(\ref{eq:ernstpot}) 
\beq
\curl (M^2 z) =-\curl (\bar\Phi \nabla \Phi)
\eeq
to replace the other term in this expansion, one finds
\begin{eqnarray}
\label{nuovosimon1}
{\rm SimonMars}
&=& M^{-3}\Scurl (M^3 \, z\otimes z)
 - \frac32 [\SYM (z\otimes (x\times \bar x)]\TF \nonumber\\
 && \quad 
-  z\times  \SYM (x \otimes \bar x ) \ .
\end{eqnarray}

In the vacuum case where $x=0$, this reduces to
\beq
\label{nuovosimon2}
{\rm SimonMars} = M^{-3}\Scurl (M^3 \, z\otimes z)\ , 
\eeq
which can be rewritten in terms of the Ernst potential by making the substitution 
$z= -\frac12 M^{-2} \nabla \mathcal{E}$
\beq
\label{simonmarsvacuum}
{\rm SimonMars}
= \frac14 M^{-3}
  \Scurl (M^{-1}\nabla \mathcal{E} \otimes \nabla \mathcal{E})
\ .
\eeq
As shown in \cite{cys2}, this is proportional to the form of the Simon tensor given in \cite{Kramer} and to the original form given by Simon. 

\section{Rigidly rotating stationary fluids}

Consider a perfect fluid at rest in the reference frame adapted to the timelike Killing direction $u$.
The energy-momentum tensore can be written as
\beq
T^{\alpha\beta}=(\rho+p)u^\alpha u^\beta +pg^{\alpha\beta}
\eeq
or in components 
\beq
T_{\top \top}= \rho\ , \quad T_{\top a}=0\ , \quad T_{ab}=ph_{ab}\ ,
\eeq
where the trace is  $T=T^\alpha{}_\alpha=3p-\rho$.
The conservation equation $\nabla_\beta T^{\alpha\beta}=0$ gives
\beq
(\rho + p) \nabla \ln M + \nabla p=0\ ,
\eeq
or assuming $p=p(f)$ with $f=M^2$
\beq
\frac{\rho + p}{2f}+ \frac{\partial p}{\partial f}=0\ .
\eeq
In this case one can immediately verify that $J^{(G)}=0$. Indeed the result follows  from Eq.~(\ref{JGfromdef}) using the fact that ${\rm Ricci}_{ab}$ is a pure trace term (killed by exterior product and Scurl operations) and $R_{\top a}$ vanishes identically.

\section{Explicit examples}

\noindent{\it The Kerr-Newman-Taub-NUT spacetime}\medskip

In  Boyer-Lindquist coordinates $(x^\alpha)=(t,r,\theta,\phi)$, with the abbreviations
$(c,s)=(\cos\theta,\sin\theta)$,
the (exterior) Kerr-Newman-Taub-NUT (KNTN)
spacetime line element is
\cite{DemNew}
\begin{eqnarray}
\label{metrica}
\rmd s^2
   &=&-\frac{1}{\Sigma}(\Delta -a^2 s^2)\rmd t^2
    +\frac{2}{\Sigma}[\Delta \chi -a(\Sigma +a \chi)s^2]\rmd t
\rmd \phi \nonumber \\ 
    &&+\frac{1}{\Sigma}[(\Sigma +a \chi)^2 s^2 -\chi^2 \Delta ]\rmd \phi^2
    +\frac{\Sigma}{\Delta}\rmd r^2 +\Sigma d\theta^2\ ,
\end{eqnarray}
and the corresponding electromagnetic Faraday tensor can be expressed in terms of the 2-form
\begin{eqnarray}
\label{max}
F&=&\frac{Q}{\Sigma^2}\{ [r^2-(\ell+ac)^2]\rmd r \wedge (\rmd t
-\chi\rmd \phi)\nonumber \\
    && +2rs (\ell+ac)\rmd \theta \wedge
[(r^2+a^2+\ell^2)\rmd \phi -a \rmd t]\}\ ,
\end{eqnarray}
which is the exterior derivative of the vector potential
\beq
A=-\frac{Qr}{\Sigma}(\rmd t -\chi \rmd \phi)\ .
\eeq 
Here  $\Sigma$, $\Delta$, and $\chi$ are defined by
\beq
\fl\qquad
\Sigma = r^2 +(\ell+ac)^2,\ 
\Delta = r^2-2\mathcal{M}r-\ell^2 +a^2+Q^2,\ 
\chi = a s^2 -2\ell c\ .
\eeq
Units are chosen so that $G=c=1$, so the parameters $(\mathcal{M},Q,a,\ell)$ all have the
dimension of length.
The source has mass $\mathcal{M}$, electric charge $Q$, angular momentum $J=\mathcal{M}a$
(i.e., gravitomagnetic dipole
moment) along the $\theta=0$ direction, and gravitomagnetic monopole moment
$\mu=-\ell$,
where $\ell$ is the NUT parameter.

The unit timelike vector aligned with the Killing direction $u=\frac{1}{M} \partial_t$ ($M^{-2}=\Sigma/(\Delta -a^2s^2)$ has nonzero acceleration and vorticity (if $a\neq0$), summarized by the complex vector
\beq
\label{eq:zZ}
z = 
\frac{\{[r-i(\ell+ac)](\mathcal{M}+i\ell)-Q^2\}}{[r+i(l+ac)]^2 (\Delta-a^2s^2)[r-i(l+ac)]} 
   (-\Delta \partial_r+ias \partial_\theta)\ ,
\eeq
while the electric and magnetic parts of the Weyl tensor are summarized by
\beq
Z =
\frac{3(\Delta-a^2s^2)}{\{[r-i(\ell+ac)](\mathcal{M}+i\ell)-Q^2\}} [z\otimes z]\TF
\ ,
\eeq
so that the Simon-Mars tensor (and hence the Simon tensor as well) vanishes.

\bigskip\noindent{\it The Kramer solution}\medskip

Using a system of coordinates $(t,f,\Omega, \phi)$, directly related to the Ernst function $\mathcal{E}=f+i\Omega$, the rigidly rotating perfect fluid solution due to Kramer (see Eq.~25 of \cite{Kramersol}, with $c=1$) is given by
\beq
\rmd s^2=-f(\rmd t -\frac1f \rmd \phi)^2+\frac1f \left(\frac{2}{\beta-F_0}\rmd f^2 +\frac{2}{\beta+F_0}\rmd \Omega^2 +\frac{\beta-F_0}{\beta+F_0}\rmd \phi^2\right),
\eeq
where $\beta\equiv \beta(f)= -F_0 +d (e^f/f)$ and $d, F_0$ are constants.
The energy-momentum tensor of the fluid is
\beq
T=[\rho(f)+p(f)]u\otimes u + p(f) g, \qquad u=\frac{1}{\sqrt{f}}\partial_t
\eeq
and the energy density and the pressure depend only on $f$ 
\beq
\rho=\frac{F_0}{4\kappa} (3f-2), \qquad p=\frac{F_0}{4\kappa} (2-f)
\eeq
and are related by the equation of state 
\beq
\rho +3 p= F_0/\kappa,
\eeq
where  $\kappa$ is the Einstein equation constant.
The four velocity of the fluid $u$ is aligned with the timelike Killing direction and has nonzero acceleration and vorticity, so that
\beq
z=-a-i\omega=-\frac{1}{2f} (\rmd f -i \rmd \omega ).
\eeq
The electric and magnetic parts of the Weyl tensor are summarized by
\begin{eqnarray}
Z&=&\frac{1}{4f}\left[\frac13 \frac{3de^f-4F_0f}{de^f-2F_0f}\rmd f \otimes \rmd f -i(\rmd f \otimes \rmd \Omega +\rmd \Omega \otimes \rmd f)
\right.\nonumber \\
&&\left.-\frac{e^{-f}}{3fd}(3de^f-2F_0f)\rmd \Omega \otimes \rmd \Omega
+\frac{e^{-f}}{3fd}(de^f-2F_0f)\rmd \omega \otimes \rmd \omega \right]\nonumber \\
&&=f\, [z\otimes z]\TF.
\end{eqnarray}
A straightforward calculation shows that the Simon-Mars (and hence the Simon tensor too) is identically zero. The Cotton current $J^{(G)}$ is also zero as shown above.

\section{Conclusions}

The gravitoelectric description of the Simon and Simon-Mars tensors which play a crucial role in characterizing the Kerr spacetime has been generalized to the class of stationary electrovacuum spacetimes and stationary rigidly rotating perfect fluid sources.
Again the alignment of the principal null directions of the Weyl and the Papapetrou is equivalent to the vanishing of the Simon-Mars tensor, which in turn forces the generalized Simon tensor to vanish as well. Examples concerning the electrovac Kerr-Newman-Taub-NUT spacetime and the perfect fluid Kramer spacetime have been presented.

\end{document}